\documentclass[a4paper,11pt]{article}
\pdfoutput=1
\usepackage{jcappub}
\usepackage{amssymb}
\usepackage{graphicx}
\usepackage{amsmath}
\usepackage{hyperref}
\usepackage{mathrsfs}
\usepackage{afterpage}

\newcommand{\dif}{\mathrm{d}}
\newcommand{\me}{\mathrm{e}}

\newcommand{\erf}{\mathrm{erf}}
\newcommand{\erfc}{\mathrm{erfc}}
\newcommand{\Ierfc}{\mathrm{Ierfc}}
\newcommand{\eV}{\mathrm{eV}}

\newcommand{\Mpc}{\mathrm{Mpc}}
\newcommand{\Gpc}{\mathrm{Gpc}}
\newcommand{\bm}{\mathbf}

\newcommand{\be}{\begin{equation}}
\newcommand{\ee}{\end{equation}}
\newcommand{\bea}{\begin{eqnarray}}
\newcommand{\eea}{\end{eqnarray}}
\newcommand{\nn}{\nonumber}

\newcommand{\bk}{\mathbf{k}}

\title{Pulsar Timing Array Constraints on the Induced Gravitational Waves}

\author[a,d]{Rong-Gen Cai,}
\author[b]{Shi Pi,}
\author[c]{Shao-Jiang Wang,}
\author[a,d,1]{Xing-Yu Yang\note{Corresponding author.} }

\affiliation[a]{CAS Key Laboratory of Theoretical Physics, Institute of Theoretical Physics, Chinese Academy of Sciences, P.O. Box 2735, Beijing 100190, China}
\affiliation[b]{Kavli Institute for the Physics and Mathematics of the Universe (WPI), The University of Tokyo Institutes for Advanced Study, The University of Tokyo, 5-1-5 Kashiwanoha, Kashiwa, Chiba 277--8583, Japan}
\affiliation[c]{Tufts Institute of Cosmology, Department of Physics and Astronomy, Tufts University, 574 Boston Avenue, Medford, Massachusetts 02155, USA}
\affiliation[d]{School of Physical Sciences, University of Chinese Academy of Sciences, No.19A Yuquan Road, Beijing 100049, China}

\emailAdd{cairg@itp.ac.cn}
\emailAdd{shi.pi@ipmu.jp}
\emailAdd{schwang@cosmos.phy.tufts.edu}
\emailAdd{yangxingyu@itp.ac.cn}

\abstract{
 If the black holes detected by LIGO/VIRGO are primordial black holes (PBHs) sourcing from a large primordial curvature perturbation on small scales, the corresponding induced gravitational waves (GWs) would peak at nanohertz that is detectable by the current and future observations of pulsar timing array (PTA). In this paper we show that with the mass function estimated from the merger rate of LIGO O1 and O2 events, the induced GWs from such a curvature perturbation with a Gaussian narrow peak at some small scale would be in a seemingly mild tension with current constraints from PTA. However, if  the curvature perturbation is of local-type non-Gaussianity with a non-linear parameter $f_\text{NL}\gtrsim\mathcal{O}(10)$, the  tension could be relieved. Nevertheless, such an induced GWs must be detectable by the Square Kilometer Array in a decade or less.
 \begin{flushleft}
 \textcolor{blue}{Preprint:	IPMU19-0095}
 \end{flushleft}
}

\begin{document}
\maketitle

\section{Introduction}\label{intro}

LIGO/VIRGO have detected gravitational waves (GWs) from mergers of binary black holes (BBHs) and binary neutron stars~\cite{Abbott:2016blz,Abbott:2016nmj,Abbott:2017vtc,Abbott:2017gyy,Abbott:2017oio,TheLIGOScientific:2017qsa}. It is possible that these black holes we observed are of the primordial origin~\cite{Bird:2016dcv,Sasaki:2016jop,Chen:2016pud,Wang:2016ana,Blinnikov:2016bxu,Ali-Haimoud:2016mbv,Garcia-Bellido:2017imq,Guo:2017njn,Zumalacarregui:2017qqd,Clesse:2016vqa,Garcia-Bellido:2017aan,Mandic:2016lcn,Raidal:2017mfl,Clesse:2016ajp}, which are formed by gravitational collapse when the curvature perturbation is of $\mathcal{O}(1)$ at the horizon reentry~\cite{Zeldovich:1967,Hawking:1971ei,Carr:1974nx,Meszaros:1974tb,Carr:1975qj}. During inflation, the primordial curvature perturbation $\mathcal {R}$ is generated by quantum fluctuations, which is then frozen when it is  stretched outside the horizon by inflation,  and finally seeds the temperature anisotropies and the large scale structure inhomogeneities on large scales  observed today. It is already well constrained by the observations of the cosmic microwave background (CMB) that $\mathcal{R}$ is nearly scale invariant and Gaussian, and has an amplitude of order $10^{-5}$ on scales roughly larger than $1\text{Mpc}$~\cite{Akrami:2018odb,Aghanim:2018eyx}. However, the curvature perturbation on small scales is unknown, because the resolution of the CMB experiments is limited, and the other observations are not so accurate. Substantial primordial black holes (PBHs) can be the seeds for galaxy formation~\cite{Nakama:2017xvq,Bean:2002kx,Kawasaki:2012kn,Carr:2018rid}, the dark matter candidate~\cite{Frampton:2009nx,Carr:2009jm,Carr:2016drx,Green:2004wb,Nakama:2019htb}, or the sources of LIGO/VIRGO detection, depending on their masses and abundance at formation, which can be constrained by PBH remnants that survive Hawking radiation, star-capture processes, microlensing, CMB $\mu$-distortion, and so on.

Such a peak in the curvature perturbation is possible in some inflation models, which has attracted much attention recently~\cite{Frampton:2010sw,Kawasaki:2012wr,Pi:2017gih,GarciaBellido:1996qt,Kawasaki:1997ju,Yokoyama:1998pt,Kohri:2012yw,Clesse:2015wea,Inomata:2017okj,Garcia-Bellido:2017mdw,Kannike:2017bxn,Inomata:2017vxo,Ando:2017veq,Ando:2018nge,Espinosa:2017sgp,Espinosa:2018eve,Cheng:2016qzb,Cheng:2018yyr,Clesse:2018ogk}. It is also well known that the large scalar perturbation can induce large tensor perturbations since they are coupled at the nonlinear order. If the curvature perturbation becomes large at small scales,  the tensor perturbation it couples to at the nonlinear order gets also large, which can evolve to today as an isotropic stochastic bacground of GWs~\cite{Matarrese:1992rp,Matarrese:1993zf,Matarrese:1997ay,Noh:2004bc,Carbone:2004iv,Nakamura:2004rm,Ananda:2006af,Baumann:2007zm,Osano:2006ew,Alabidi:2012ex,Alabidi:2013wtp,Assadullahi:2009nf,Inomata:2016rbd,Orlofsky:2016vbd,Kohri:2018awv,Assadullahi:2009jc,Biagetti:2014asa,Gong:2017qlj,Giovannini:2010tk,Cai:2018dig,Unal:2018yaa,Bartolo:2018rku,Byrnes:2018txb,Inomata:2018epa,Dalianis:2018ymb,Cai:2019amo,Cai:2018tuh,Cai:2019jah,DeLuca:2019qsy,Tada:2019amh,Inomata:2019zqy,Inomata:2019ivs,Xu:2019bdp}. The peak frequency of the GW spectrum is solely determined by the peak scale of the curvature perturbation spectrum, which is connected to the PBH mass generated from this peak by
\begin{equation} 
f_\text{peak}\sim6.7\times10^{-9}\left(\frac{M_\text{PBH}}{M_\odot}\right)^{-1/2}~\text{Hz}. 
\end{equation}
If the black holes detected by LIGO/VIRGO are primordial, we must have such accompanied stochastic background of GWs induced by such scalar perturbations, which peaks around $\text{nHz}$. The peak amplitude of the spectrum for the stochastic GW is determined by the peak value of the scalar power spectrum that determines the PBH abundance, which in turn can be constrained by the black hole merger rate from LIGO/VIRGO events. 

Millisecond pulsars have very stable rotation periods, thus the pulses we observe can be used as a tool to detect the spacetime metric between the observer and pulsars. By studying the time of arrivals from many pulsars in the sky, we can extract the information of the stochastic GWs from this pulsar timing array (PTA)~\cite{Sazhin:1977tq,Detweiler:1979wn}. Currently, there are three PTA projects: the European Pulsar Timing Array (EPTA \cite{Desvignes:2016yex}), the Parkes Pulsar Timing Array (PPTA \cite{Hobbs:2013aka}), and the North American Observatory for Gravitational Waves (NANOGrav \cite{McLaughlin:2013ira}), while the International Pulsar Timing Array (IPTA \cite{Verbiest:2016vem}) is the combination of them. The GW frequency range is set by the total observation time, which is roughly $f\sim1/T_\text{tot}\sim10^{-9}~\text{to}~10^{-7}~\mathrm{Hz}$, right in the range for the induced GWs accompanied with the solar-mass PBHs.
For now there is no evidence of stochastic GW background. For instance, 2015 EPTA data sets an upper bound $\Omega_\text{GW}h^2<1.1\times10^{-9}$ at $2.8~\text{nHz}$~\cite{Lentati:2015qwp} at $95\%$ confidence level. However, this is in conflict with what the LIGO detection has predicted, if the black holes detected by LIGO/VIRGO are primordial~\cite{Saito:2008jc,Saito:2009jt}.

Previous works assume that the scalar perturbation is Gaussian and has a narrow peak on a specific small scale. It is known that a broad peak can make the conflict even worse~\cite{Bugaev:2009zh,Bugaev:2010bb}. However, introducing non-Gaussianity could be helpful. In the inflation models which predict a peak in the power spectrum of the curvature perturbation, it is quite natural to have non-negligible non-Gaussianity. The nonlinear relation between density contrast and the curvature perturbation provides another source of non-Gaussianity. As PBHs are produced at the large amplitude tail of the probability distribution of the curvature perturbation,
positive non-negligible non-Gaussianity can enhance the abundance of the PBH formation. Equivalently, given the PBH abundance, a non-Gaussian scalar perturbation may induce smaller GWs, which may escape from the current PTA constraint. Ref. \cite{Nakama:2016gzw} has considered non-Gaussian scalar perturbations. By choosing a monochromatic mass function of PBH and an order estimation for the induced GWs, they found that the PTA constraint can be evaded if non-Gaussianity is large enough. 

In this paper, we study this possibility carefully by reconstructing the mass function from the parameters of the  LIGO events in O1 and O2, following the method established in \cite{Raidal:2017mfl} and calculating the GWs induced by the non-Gaussian scalar perturbations by the formula obtained previously in \cite{Cai:2018dig}.
With the EPTA constraints, we found that it is still possible for the black holes detected by LIGO/VIRGO to be primordial if $f_\text{NL}\gtrsim\mathcal{O}(10)$.

This paper is organized as follows:
in Section \ref{sec:PBHandGW}, we review the derivations of mass function of PBHs and energy-density spectrum of induced GWs;
in Section \ref{sec:Gaussian}, we calculate the mass function of PBHs which originate from Gaussian primordial curvature perturbations and find that it can be well parametrized;
in Section \ref{sec:fit}, we make a maximum likelihood analysis to find the best-fit parameters by a fitting formula, and confirm the corresponding induced GW is incompatible with the constraints from PTA experiments;
in Section \ref{sec:NG}, we discuss how to evade the conflict by considering the primordial curvature perturbations  with local-type non-Gaussianity.
We conclude in Section \ref{sec:con}.

\section{Mass function of PBHs and Induced GWs}\label{sec:PBHandGW}

In this section we review the PBH mass function and its associated induced GWs. 
Assuming the peak of the power spectrum for the curvature perturbation reenters the Hubble horizon in the radiation dominated era, PBH will form soon after this reentry time. The PBH mass fraction  at formation  $\beta_{m_{\mathrm{H}}}$ of horizon mass $m_{\mathrm{H}}$ is usually calculated by the Press-Schechter approach~\cite{Young:2014ana,Byrnes:2018clq,Wang:2019kaf}, which is given by\footnote{For an alternative approach based on the peak theory, see ~\cite{Yoo:2018esr,Yoo:2019pma}}
\begin{equation}\label{eq:beta}
\beta_{m_{\mathrm{H}}}= 2 \int_{\Delta_{c}}^{\infty} \frac{m}{m_{\mathrm{H}}} P(\Delta) d\Delta,
\end{equation}
where $\Delta=(\delta\rho-\rho)/\rho$ is the density contrast, $P(\Delta)$ is its probability distribution function, and $\Delta_{c}$ is the critical density contrast above which PBHs will form.
Assuming $P(\Delta)$ is Gaussian,
\begin{equation}
P(\Delta) = \frac{1}{\sqrt{2\pi\sigma^{2}}} \exp \left( -\frac{\Delta^{2}}{2\sigma^{2}} \right),
\end{equation}
where $\sigma^{2}$ is the variance of density contrast given by integrating the density power spectrum $\mathcal{P}_{\Delta}(\eta, q)$,
\begin{equation}
\sigma^{2} = \int_{0}^{\infty} \frac{dq}{q} \tilde{W}^{2}(R, q) \mathcal{P}_{\Delta}(\eta, q),
\end{equation}
where $\tilde{W}(R, q)=\exp(-q^{2}R^{2}/2)$ is Fourier transform of a volume-normalised Gaussian window smoothing function, $R$ is horizon scale at a given conformal time $\eta$. For the subtlety of choosing the window function, see~\cite{Ando:2018qdb,Young:2019osy}.
Using the relation between curvature perturbation $\mathcal{R}$ and density contrast $\Delta$,
\begin{equation}
\Delta(t,q)=\frac{2(1+w)}{5+3w} \left( \frac{q}{aH} \right)^{2} \mathcal{R}(q),
\end{equation}
where $w$ is the equation-of-state parameter, which is $1/3$ in the radiation dominated era, $\sigma^{2}$ can be written as
\begin{equation}
\sigma^{2} = \int_{0}^{\infty} \frac{dq}{q} \tilde{W}^{2}(R, q) \left( \frac{4}{9} \right)^{2} \left( \frac{q}{aH} \right)^{4} \mathcal{P}_{\mathcal{R}}(\eta, q),
\end{equation}
where $\mathcal{P}_{\mathcal{R}}(\eta, q)$ is the power spectrum of curvature perturbations.

For a wavenumber $k$ corresponding to horizon reentry in the radiation dominated era, we have $k=\eta^{-1}=aH=\mathcal{H}$, $R=\mathcal{H}^{-1}=k^{-1}$, and then
\begin{equation}\label{def:sigma}
\sigma^{2}(k) =  \int_{0}^{\infty} \frac{dq}{q} \tilde{W}^{2}(R=k^{-1}, q) \left( \frac{4}{9} \right)^{2} \left( \frac{q}{k} \right)^{4} T^{2}(\eta=k^{-1}, q) \mathcal{P}_{\mathcal{R}}(q),
\end{equation}
where $T(\eta, q)=3(\sin y - y \cos y)/y^{3}$ with $y \equiv q\eta/ \sqrt{3}$ is the transfer function and $\mathcal{P}_{\mathcal{R}}(q)$ is the power spectrum of primordial curvature perturbations.

It is worth noting that the energy density of PBHs evolves like matter during radiation dominated era thus increases proportional to $a$ until equality, and $\beta_{m_{\mathrm{H}}}$ represents the fraction of PBHs when horizon mass is $m_{\mathrm{H}}$, thus at equality, the fraction $\beta_{\mathrm{eq}}$ is
\begin{equation}
\beta_{\mathrm{eq}} = \frac{a_{\mathrm{eq}}}{a_{m_{\mathrm{H}}}} \beta_{m_{\mathrm{H}}},
\end{equation}
where $a_{\mathrm{eq}}$ is the scale factor at equality and $a_{m_{\mathrm{H}}}$ is the scale factor when horizon mass is $m_{\mathrm{H}}$.
Integrating $\beta_{\mathrm{eq}}$ over all possible PBH formation time, we obtain the total abundance of PBHs at equality,
\begin{equation}
\Omega_{\mathrm{PBH,eq}} = \int_{-\infty}^{\infty} d \ln m_{\mathrm{H}}\ \frac{a_{\mathrm{eq}}}{a_{m_{\mathrm{H}}}} \beta_{m_{\mathrm{H}}}.\\
\end{equation}

The mass function of PBHs $f(m)$ is defined as the fraction of cold dark matter (CDM) made up of PBHs with mass $m$,
\begin{equation}
f(m) \equiv \frac{1}{\Omega_{\mathrm{CDM}}} \frac{d \Omega_{\mathrm{PBH}}}{d \ln m}=\frac{\Omega_{m}}{\Omega_{\mathrm{CDM}}} \frac{d \Omega_{\mathrm{PBH,eq}}}{d \ln m},
\end{equation}
where $\Omega_{\mathrm{CDM}}/\Omega_{m}\approx 0.84$ \cite{Aghanim:2018eyx}. The total fraction of PBHs in CDM is
\begin{equation}
    f_{\mathrm{PBH}} \equiv \frac{\Omega_{\mathrm{PBH}}}{\Omega_{\mathrm{CDM}}} = \int_{-\infty}^{\infty} f(m) d\ln m.
\end{equation}
The relation of the PBH mass and the density contrast is given by the critical collapse~\cite{Yokoyama:1998xd,Niemeyer:1999ak}
\begin{equation}
m = m_{\mathrm{H}} K (\Delta -\Delta_{c})^{\gamma},
\end{equation}
where the constants $K=3.3$, $\gamma=0.36$, $\Delta_{c}=0.45$ during radiation-dominated era are given by numerical studies.
Therefore, the mass function of PBHs can be written as
\begin{equation}\label{eq:fm}
\begin{aligned}
f(m) 
&= 2 \frac{\Omega_{m}}{\Omega_{\mathrm{CDM}}} \int_{-\infty}^{\infty} d \ln m_{\mathrm{H}}\ \frac{a_{\mathrm{eq}}}{a_{m_{\mathrm{H}}}} \frac{m}{m_{\mathrm{H}}} P(\Delta) \frac{d\Delta}{d \ln m},
\end{aligned}
\end{equation}
with
\begin{equation}
\Delta = \left( \frac{m}{K m_{\mathrm{H}}} \right)^{1/\gamma} + \Delta_{c}.
\end{equation}

Another important phenomenon generated by the peak of the power spectrum for the curvature perturbation is the induced GWs. 
We will shortly review it here for completeness following \cite{Cai:2019amo,Cai:2019cdl}. The perturbed metric in the conformal Newton gauge reads,
\begin{equation}
    ds ^{2} =a ^{2} ( \eta ) \left\{ - ( 1+2 \Phi ) d\eta ^{2} + \left[ ( 1-2\Phi ) \delta _{ij} + \frac{1}{2} h _{ij} \right] dx ^{i} dx ^{j} \right\},
\end{equation}
where $ \eta $ is the conformal time, $ \Phi $ is the Newton potential and $ h _{ij} $ is the tensor mode of the metric perturbation in the transverse-traceless gauge. As we neglect the anisotropic stress tensor, the scalar perturbations in time and diagonal spatial components are identical. 
$h_{ij}(\eta,\mathbf{x})$ can be expanded as follows,
\begin{equation}
h_{ij}(\eta,\mathbf{x})=\int\frac{d^3k}{(2\pi)^{3/2}}\sum_{\lambda=+,\times}e_{ij}^\lambda(\hat k)h_{\mathbf{k},\lambda}(\eta)e^{i\mathbf{k}\cdot\mathbf{x}},
\end{equation}
where $+$ and $\times$ are the two polarizations, respectively. 
The equation of motion for $ h_{ij} $ can be derived from the perturbed Einstein equation up to the second order, which in the momentum space can be written as
\begin{equation}\label{eq:eof}
    h_\mathbf{k} ''+2 \mathcal{H} h_\mathbf{k} '+k^{2} h_\mathbf{k} =2\mathscr{P}_{ij}^{lm}e^{ij}T_{lm}(\bm{k}, \eta ) ,
\end{equation}
for each polarization, where $\mathcal{H}$ is the conformal Hubble parameter, $\mathscr{P}_{ij}^{lm}$ is the projection operator to the transverse-traceless part, and a prime denotes the derivative with respect to $\eta$. The source term $T_{lm}( \bm{k} ,\eta )$ is of second order in scalar perturbations, 
\begin{equation}
T_{lm}=-2\Phi\partial_l\partial_m\Phi+\partial_l\left(\Phi+\mathcal{H}^{-1}\Phi'\right)\partial_m\left(\Phi+\mathcal{H}^{-1}\Phi'\right).
\end{equation}
This equation can be solved with Green's function method in the momentum space, and its solution is
\be
h_{\bk}(\eta)=\frac2{a(\eta)}\int^\infty_0d\tilde\eta~a(\tilde\eta)G_k(\eta;\tilde\eta)\mathscr{P}^{lm}_{ij}(\hat k)e^{ij}T_{lm}(\tilde\eta,\bk).
\ee
The two-point function of $h_\bk$ can be written as
\begin{equation}\label{eq:power_h}
    \langle h _{\bm{k}} ( \eta ) h _{\bm{l} } ( \eta ) \rangle = \delta ^{( 3 )}( \bm{k} + \bm{l} ) \frac{2 \pi ^{2}}{k ^{3}} \mathcal{P} _{h} ( \eta,k ) .
\end{equation}
It is usual to describe the stochastic GWs by its energy density per logarithmic frequency interval normalized by the critical density,
\begin{equation}
    \Omega _{\mathrm{GW}} ( \eta ,k ) = \frac{1}{24} \left( \frac{k}{\mathcal{H ( \eta )}} \right) ^{2} \overline{\mathcal{P} _{h} ( \eta, k )},
\end{equation}
where the two polarization modes of GWs have been summed over, and the overline denotes average over a few wavelengths. Using the relation between scalar perturbations $\Phi$ and curvature perturbations $\mathcal{R}$ in the radiation-dominated era, $\Phi=-(2/3)\mathcal{R}$, we can calculate the GW spectrum induced by the curvature perturbation during radiation-dominated era as
\begin{equation}\label{eq:spectrum_GW}
\Omega _{\mathrm{GW}} ( k, \eta)= \frac{1}{54} \int _{0}^{\infty} \dif v \int _{|1-v|}^{1+v} \dif u~ \mathcal{T} ( u,v ) \mathcal{P} _{\mathcal{R}} ( ku ) \mathcal{P} _{\mathcal{R}} ( kv ) ,
\end{equation}
with
\begin{equation}\nn
    \begin{aligned}
        \mathcal{T} ( u,v )
&=  \frac{1}{4} \left( \frac{4v ^{2} - ( 1+v ^{2} -u ^{2} ) ^{2}}{4uv} \right) ^{2} \left[  \frac{27}{4 u ^{3} v ^{3}} ( u ^{2} +v ^{2} -3 ) \right] ^{2} \\
& \times \frac{1}{2} \left\{ \left[ -4uv+( u ^{2} +v ^{2} -3 ) \ln \left|\frac{3- ( u+v ) ^{2}}{3- ( u-v ) ^{2}} \right| \right] ^{2}  +  \left[ \pi  ( u ^{2} +v ^{2} -3 ) \Theta _{\frac{1}{2}}  ( u+v- \sqrt{3} ) \right] ^{2} \right\}.
    \end{aligned}
\end{equation}

\section{A fitting formula for the PBH mass function}\label{sec:Gaussian}

The PBHs will form when the rare peaks of the curvature perturbations reenter the Hubble horizon in the radiation dominated era.
However, the amplitude of the curvature perturbations on CMB scales are highly constrained, which makes the PBH abundance completely negligible if the curvature perturbations on small scales are extropolated linearly by its red-tilted power spectrum on CMB scales.
If the BBHs detected by LIGO are composed of PBHs, we need substantial formation of such PBHs, which requires a relatively large amplitude of the curvature perturbations of $\mathcal{O}(1)$ on the corresponding small scales.
This is consistent with the power spetrum on CMB scales if we also consider the running of running from the current observational constraints~\cite{Green:2018akb}.
For the sake of convenience, such a peaked power spectrum can be described by a narrow Gaussian peak with a characteristic wavenumber $k_*$ and a width $\sigma_{*} \ll k_{*}$,
\begin{equation}\label{eq:curvature_spectrum}
\mathcal{P}_{\mathcal{R}}(k)= \frac{\mathcal{A}_{\mathcal{R}} k_{*}}{\sqrt{2\pi}\sigma_{*}}\exp \left[ - \frac{(k-k_{*})^{2}}{2\sigma_{*}^{2}} \right],
\end{equation}
where the near-scale-invariant part from the CMB scales is neglected.

\begin{figure}[htpb]
\centering
\includegraphics[width=0.48\textwidth]{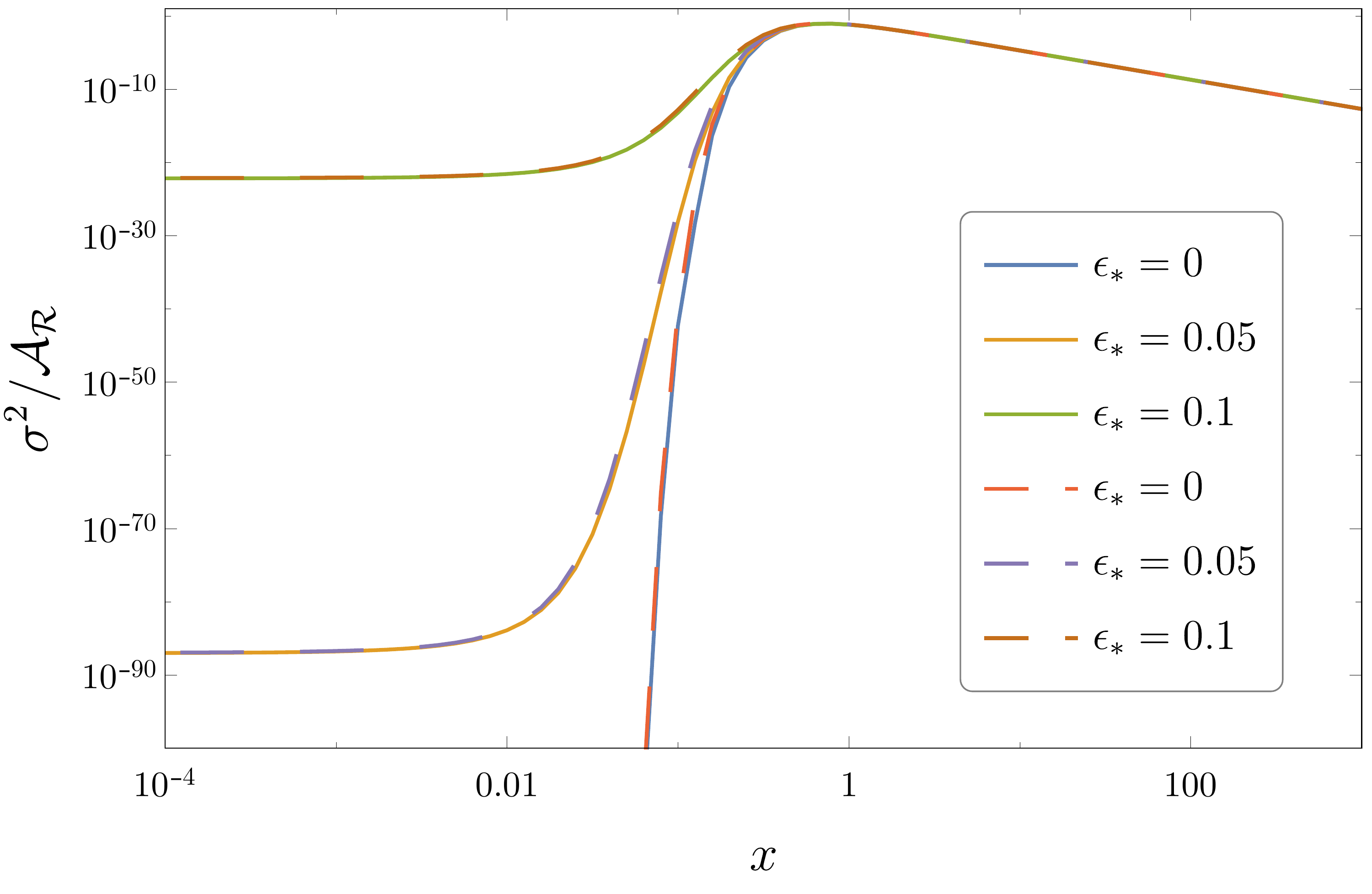}
\includegraphics[width=0.48\textwidth]{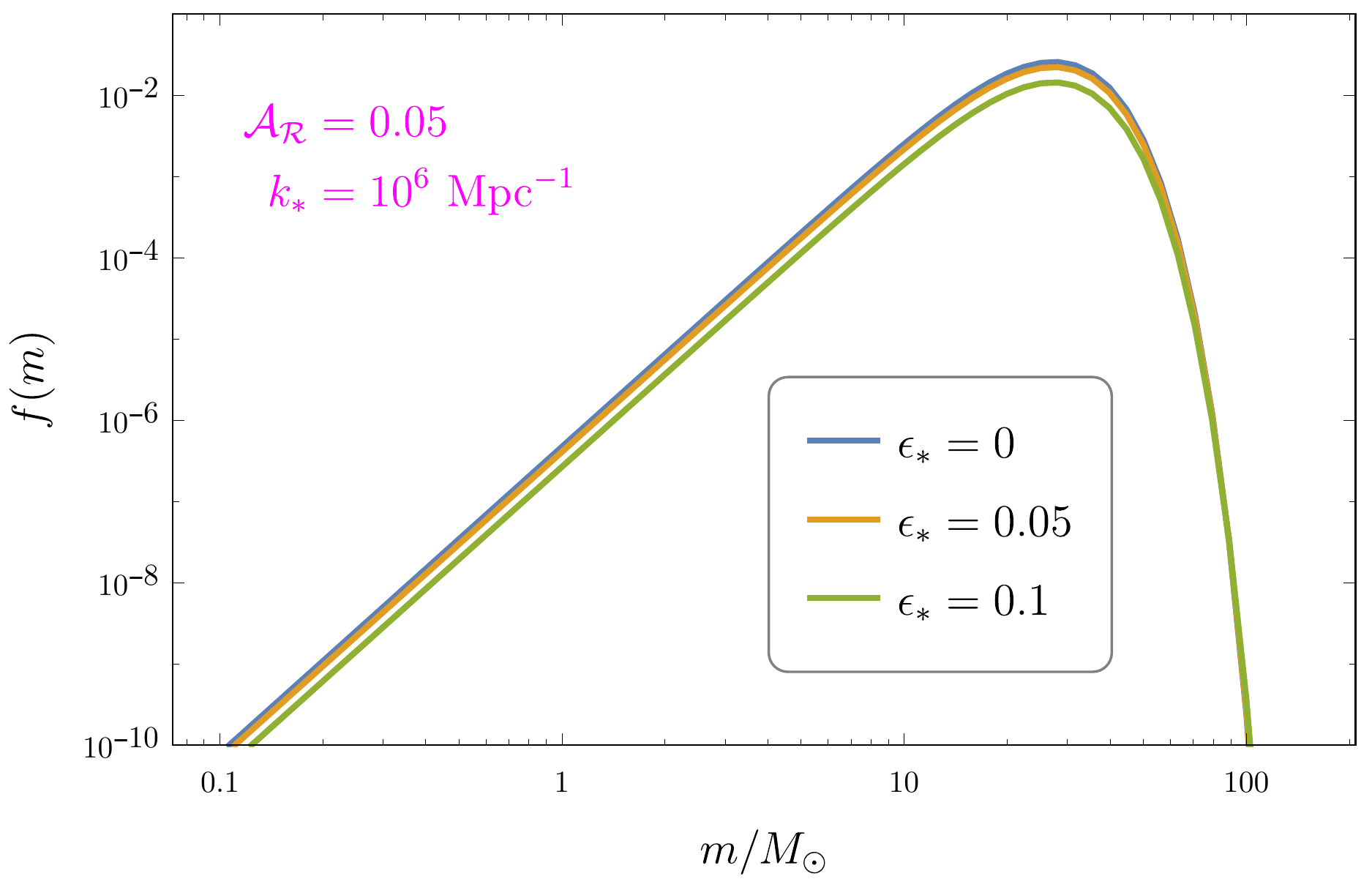}
\caption{Variance of density contrast $\sigma^{2}$ and mass function of PBHs $f(m)$ originated from power spectrum with narrow Gaussian peak  $\mathcal{P}_{\mathcal{R}}(k)$.
\textit{Left}: the exact $\sigma^{2}$ (Eq.\eqref{eq:sigma2e} ) and approximate $\sigma^{2}$ (Eq.\eqref{eq:sigma2a}) are shown by solid and dashed lines respectively, from bottom to top, $\epsilon_{*}=0, 0.05, 0.1$.
\textit{Right}: the mass function of PBHs $f(m)$ with $\mathcal{A}_{\mathcal{R}}=0.05$, $k_{*}=10^{6}\Mpc^{-1}$, from top to bottom, $\epsilon_{*}=0,0.05,0.1$, respectively.}
\label{fig:sigma_fm}
\end{figure}

By using \eqref{def:sigma}, we can calculate the variance of density contrast,
\begin{equation}\label{eq:sigma2e}
\sigma^{2}(k)= \frac{16 \mathcal{A}_{\mathcal{R}}}{\sqrt{2\pi}\epsilon_{*}} \int_{0}^{\infty} \frac{dz}{z} \exp (-z^{2}) \left[ \frac{\sqrt{3}}{z} \sin (\frac{z}{\sqrt{3}}) - \cos (\frac{z}{\sqrt{3}}) \right]^{2} \exp \left[ - \frac{(xz-1)^{2}}{2\epsilon_{*}^{2}} \right],
\end{equation}
where $\epsilon_{*} \equiv \sigma_{*}/k_{*}$, $x\equiv k/k_{*}$, $z\equiv q/k$. As the window function will greatly suppress the contribution from the superhorizon modes of $z\gtrsim1$, the main contribution of the integral comes from inside the horizon, which means that we can expand the integrand in \eqref{eq:sigma2e} in the small $z$ limit. This gives rise to
\begin{equation}
\begin{aligned}\label{eq:sigma2a}
\sigma^{2}(k) \approx &  \mathcal{A}_{\mathcal{R}} \frac{4 \sqrt{\frac{2}{\pi }}\ \me^{-\frac{1}{2 \epsilon_*^2}}}{81 \left(x^2+2 \epsilon_*^2\right)^{7/2}} \Biggl\{ -2 \epsilon_*^3 \left(x^2+2 \epsilon_*^2\right)^{3/2}  +2 \sqrt{2 \pi } x \ \me^{\frac{x^2}{2 x^2 \epsilon_*^2+4 \epsilon_*^4}} \left(x^2 \left(3 \epsilon_*^2+1\right)+6 \epsilon_*^4\right) \\
&-\sqrt{2 \pi } x \ \me^{\frac{x^2}{2 x^2 \epsilon_*^2+4 \epsilon_*^4}} \left(x^2 \left(3 \epsilon_*^2+1\right)+6 \epsilon_*^4\right) Q\left(-\frac{1}{2},\frac{x^2}{2 x^2 \epsilon_*^2 + 4 \epsilon_*^4}\right)   \Biggr\},
\end{aligned}
\end{equation}
where $Q(a,z)$ is the regularized incomplete Gamma function, defined as
\begin{equation}
Q(a,z) \equiv \frac{\Gamma(a,z)}{\Gamma(a)},\quad \Gamma(a,z) \equiv \int_{z}^{\infty} t^{a-1} \me^{-t} dt,\quad \Gamma(a) \equiv \int_{0}^{\infty} t^{a-1} \me^{-t} dt.
\end{equation}
In the left panel of Fig. \ref{fig:sigma_fm}, the exact $\sigma^{2}$ from Eq.\eqref{eq:sigma2e} and approximate $\sigma^{2}$ from Eq.\eqref{eq:sigma2a} with $\epsilon_{*}=0, 0.05, 0.1$ are shown in solid and dashed lines, respectively, which justifies our approximation.

\begin{figure}[htpb]
    \centering
    \includegraphics[width=0.9\textwidth]{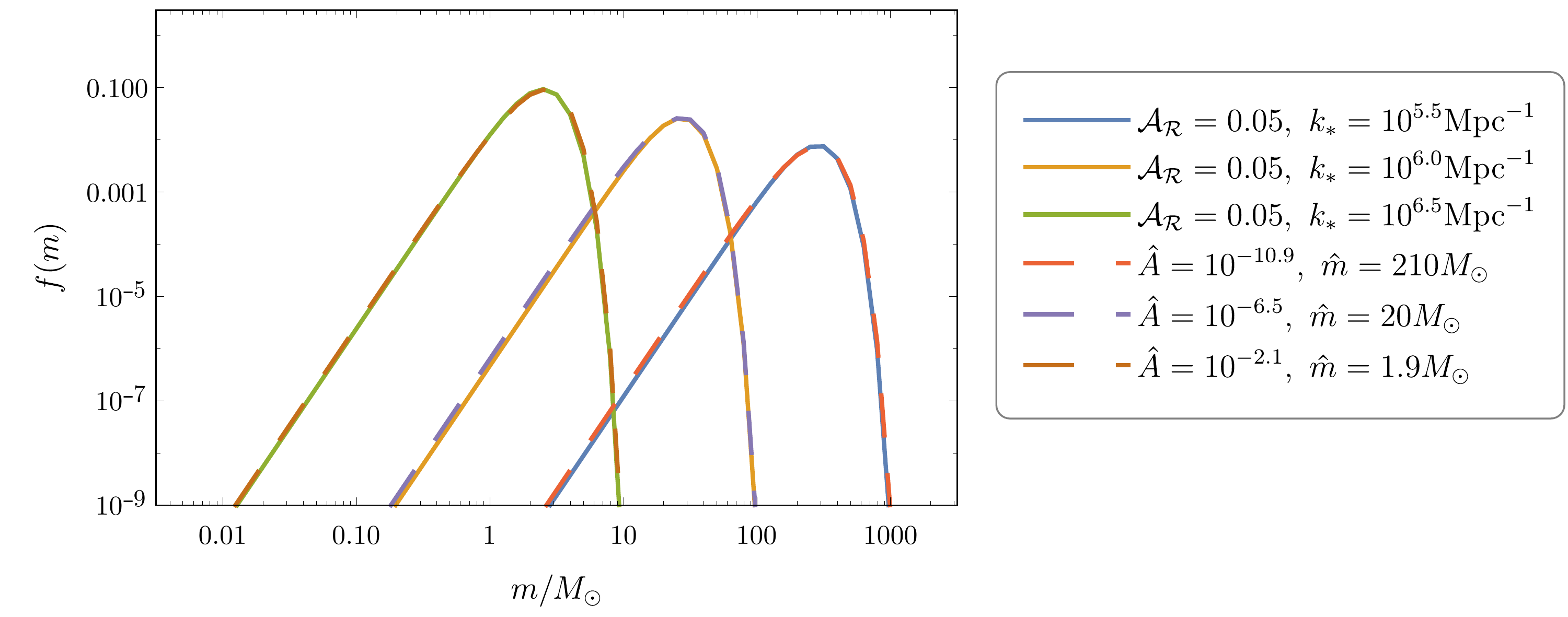}
    \includegraphics[width=0.9\textwidth]{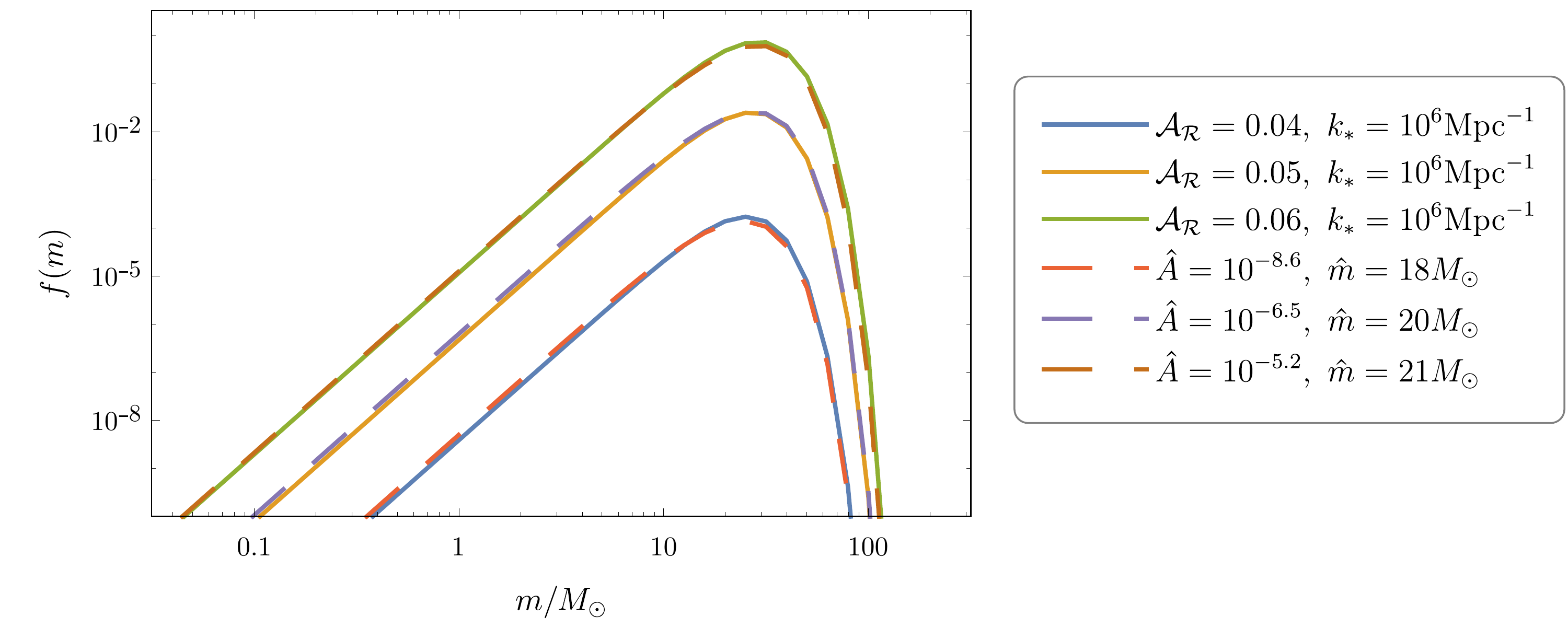}
    \caption{The exact and corresponding parametrized mass functions $f(m)$ of PBHs are shown with solid and dashed lines respectively.
    \textit{Top}: $\mathcal{A}_{\mathcal{R}}=0.05$ , from right to left,  $k_{*}=10^{5.5},10^{6.0},10^{6.5} \Mpc^{-1}$, respectively.
\textit{Bottom}: $k_{*}=10^{6} \Mpc^{-1}$, from bottom to top, $\mathcal{A}_{\mathcal{R}}=0.04,0.05,0.06$, respectively. }
    \label{fig:fmeap}
\end{figure}

In order to calculate the mass function of PBHs by Eq.\eqref{eq:fm}, one needs to know $a_{\mathrm{eq}}/a_{m_{\mathrm{H}}}$, and the relation between the wavenumber corresponding to horizon reentry $k$ and the horizon mass $m_{\mathrm{H}}$.
By definition, the horizon mass is
\begin{equation}
    m_{\mathrm{H}}=\frac{4\pi}{3} \rho H^{-3}.
\end{equation}
We can then use the 
\begin{equation}
g_{*s}(T)\ T^{3} a^{3} = \mathrm{constant},
\end{equation}
and the Friedmann equation in radiation dominated era
\begin{equation}
\frac{3}{8\pi} H^{2} = \rho \approx  \rho_{r} = \frac{\pi^{2}}{30} g_{*r}(T)\ T^{4},
\end{equation}
where $g_{*s}$ is the effective degrees of freedom for the entropy density and $g_{*r}$ is the effective degrees of freedom for relativistic particles. Then we get
\begin{equation}
\frac{a_{\mathrm{eq}}}{a_{m_{\mathrm{H}}}} = \left( \frac{g_{*s}(T_{m_{\mathrm{H}}})}{g_{*s}(T_{\mathrm{eq}})} \right)^{1/3} \frac{T_{m_{\mathrm{H}}}}{T_{\mathrm{eq}}} = \left( \frac{g_{*s}(T_{m_{\mathrm{H}}})}{3.91} \right)^{1/3} \frac{T_{m_{\mathrm{H}}}}{0.8\eV} ,
\end{equation}
and
\begin{align}
\frac{m_{\mathrm{H}}}{M_{\odot}} 
&\approx 19.1 \left( \frac{g_{*r}(T_{m_{\mathrm{H}}})}{17.77} \right)^{1/2} \left( \frac{g_{*s}(T_{m_{\mathrm{H}}})}{17.35} \right)^{-2/3}\left( \frac{k}{10^{6} \Mpc^{-1}} \right)^{-2},
\end{align}
where $T_{m_{\mathrm{H}}}$ is temperature when horizon mass is $m_{\mathrm{H}}$, and dependence of the effective degree of freedom on the  temperature is determined by the thermal history of the universe \cite{Saikawa:2018rcs,Aghanim:2018eyx,Inomata:2018epa,Ando:2018qdb}. Therefore, the mass function of PBHs can be computed as
\begin{equation}\label{eq:fme}
\begin{aligned}
f(m) = 2 \frac{\Omega_{m}}{\Omega_{\mathrm{CDM}}} \int_{-\infty}^{\infty} d \ln m_{\mathrm{H}}\ \Biggl\{ & \left( \frac{g_{*s}(T_{m_{\mathrm{H}}})}{g_{*s}(T_{\mathrm{eq}})} \right)^{1/3} \frac{T_{m_{\mathrm{H}}}}{T_{\mathrm{eq}}}  \frac{K}{\gamma\sqrt{2\pi\sigma^{2}(m_{\mathrm{H}})}} \left( \frac{m}{K m_{\mathrm{H}}} \right)^{1/\gamma+1} \\
& \times \exp \left( -\frac{1}{2\sigma^{2}(m_{\mathrm{H}})} \left[ \left( \frac{m}{K m_{\mathrm{H}}} \right)^{1/\gamma} + \Delta_{c} \right]^{2} \right) \Biggr\}.
\end{aligned}
\end{equation}
In the right panel of figure \ref{fig:sigma_fm}, the mass functions of PBHs $f(m)$ with $\mathcal{A}_{\mathcal{R}}=0.05$, $k_{*}=10^{6}\Mpc^{-1}$, $\epsilon_{*}=0,0.05,0.1$ are shown from top to bottom, respectively, which tells us that the effect of $\epsilon_{*}$ to $f(m)$ is negligible once $\epsilon_{*} \ll 1$.
Recalling that we study the case that $\sigma_{*} \ll k_{*}$ ($\epsilon_{*} \ll 1$), taking $\epsilon_{*}=0.05$ for concreteness, there are only two free parameters $\mathcal{A}_{\mathcal{R}}$ and $k_{*}$ in $f(m)$. In this case, we find a fitting formula for $f(m)$ as 
\begin{equation}\label{eq:fmap}
    f(m) \approx 2 \hat{A}(\mathcal{A}_{\mathcal{R}}, k_{*}) \left( \frac{m}{M_{\odot}} \right)^{1/\gamma+1} \exp \left( - \frac{m^{2}}{\hat{m}^{2}(\mathcal{A}_{\mathcal{R}}, k_{*})} \right),
\end{equation}
and thus
\begin{equation}
    f_{\mathrm{PBH}}= \hat{A} \left( \frac{\hat{m}}{M_{\odot}} \right)^{1/\gamma+1} \Gamma(\frac{1+\gamma}{2\gamma}).
\end{equation}
The exact $f(m)$ (Eq.\eqref{eq:fme}) can be fitted by the corresponding parametrized $f(m)$ (Eq.\eqref{eq:fmap}) with $\mathcal{A}_{\mathcal{R}}=0.05,\ k_{*}=10^{5.5},10^{6.0},10^{6.5} \Mpc^{-1}$ and $\mathcal{A}_{\mathcal{R}}=0.04,~0.05,~0.06$, $k_{*}=10^{6} \Mpc^{-1}$ with high precision as shown in figure \ref{fig:fmeap}. We can freely convert $\mathcal{A}_{\mathcal{R}}$ and $k_{*}$ into $\hat{A}$ and $\hat{m}$, and vice versa.

\section{Best-fit parameters from the observed merger rate}\label{sec:fit}

In the previous discussion, we assumed that the peak of the power spectrum for the curvature perturbation reenters Hubble horizon in the radiation dominated era. After their formation, PBHs are dynamically coupled to cosmic expansion. However, as the background energy density decays faster than that of the PBHs, the local density may become greater than ambient energy density, thus the PBHs decouple from the background evolution of the universe and  form a gravitational bound state.  The closest PBHs fall toward to each other, yet with the torques from a nearby third PBH as well as some other matter inhomogeneities, the head-on collision may be replaced by the formation of a binary of PBHs~\cite{Nakamura:1997sm}.

Taking into account the torques caused by the surrounding PBHs and linear density perturbations, the merger rate of PBH binary reads \cite{Ali-Haimoud:2017rtz,Raidal:2018bbj,Liu:2018ess}
\begin{equation}
    dR = S \frac{1.6 \times 10^{6}}{\Gpc^{3}\mathrm{yr}} f_{\mathrm{PBH}}^{\frac{53}{37}} \mu^{-\frac{34}{37}} \left( \frac{M}{M_{\odot}} \right)^{-\frac{32}{37}} \left( \frac{t(z)}{t_{0}} \right)^{-\frac{34}{37}} \tilde{f}(m_{1}) \tilde{f}(m_{2}) dm_{1} dm_{2},
\end{equation}
where $M \equiv m_{1}+m_{2}$, $\mu \equiv m_{1} m_{2}/M^{2}$. We also define the normalized mass function of PBHs
\begin{equation}
    \tilde{f}(m) \equiv \frac{f(m)}{\int_{0}^{\infty} dm f(m)}= \frac{2}{\Gamma(1+\frac{1}{2\gamma}) \hat{m}^{1/\gamma+2}} \ m^{1/\gamma+1} \exp \left( - \frac{m^{2}}{\hat{m}^{2}} \right),
\end{equation}
and the suppression factor
\begin{equation}
    S= \left( 1+\left( \frac{\Omega_{m}}{\Omega_{\mathrm{CDM}}} \frac{\sigma_{\mathrm{eq}}}{f_{\mathrm{PBH}}} \right)^{2} \right)^{-\frac{21}{74}},
\end{equation}
where $\sigma_{\mathrm{eq}} \approx 0.005$ is the variance of the density perturbations of the ambient fluid of the universe at equality. 

\begin{figure}[htpb]
    \centering
    \includegraphics[width=0.48\textwidth]{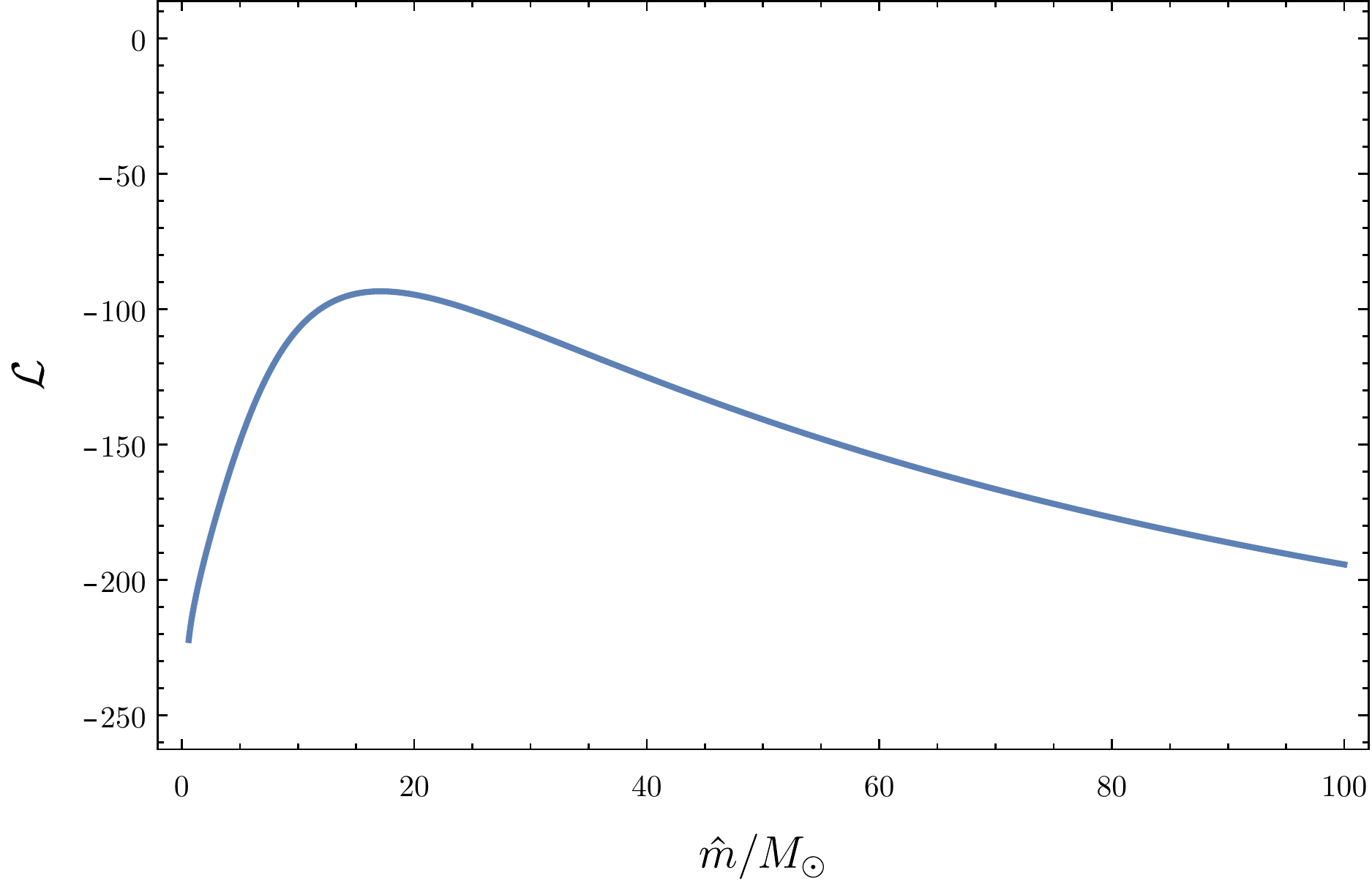}
    \includegraphics[width=0.48\textwidth]{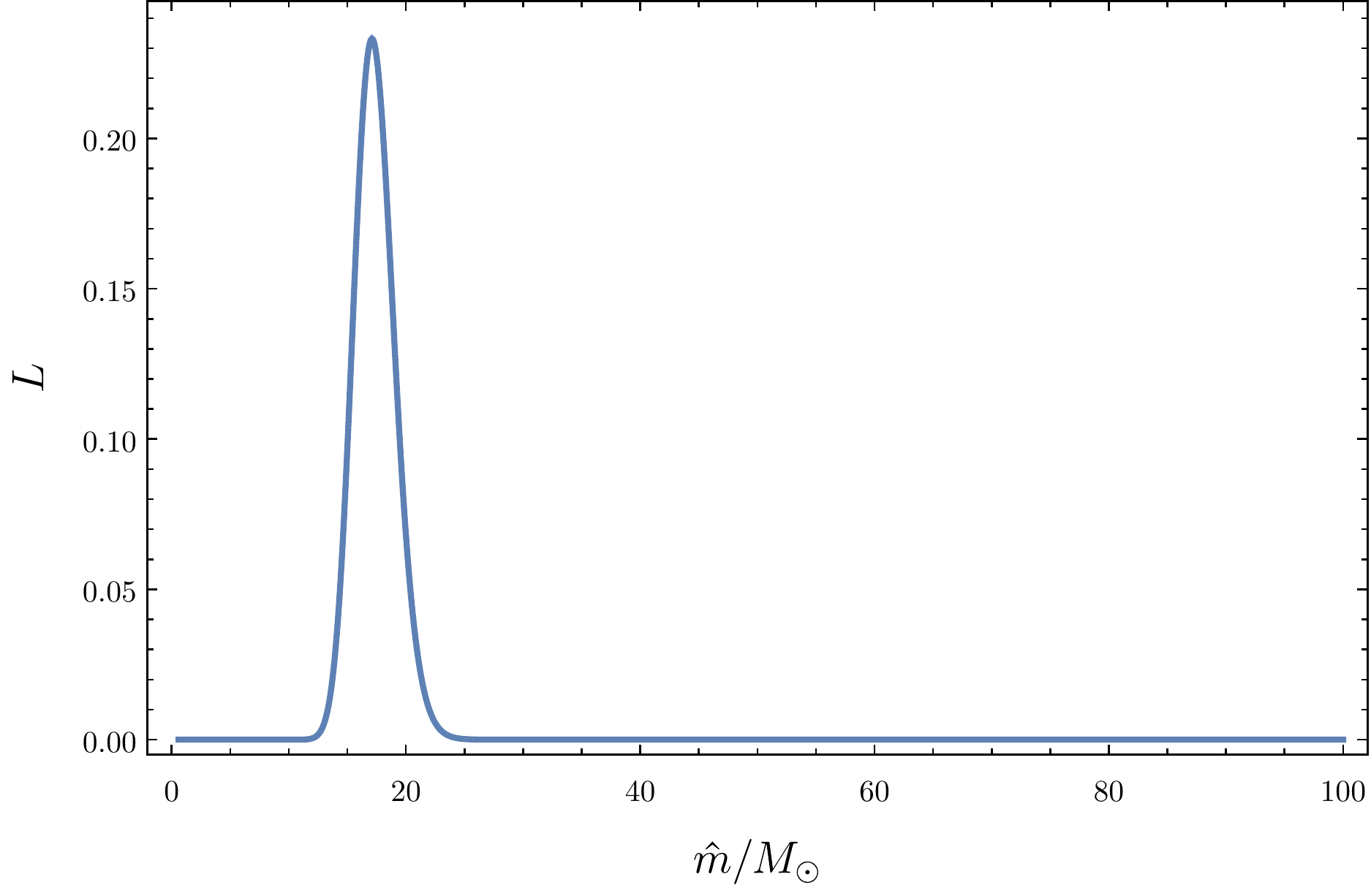}
    \caption{Log-likelihood $\mathcal{L}$ and normalized likelihood $L$.}
    \label{fig:logl}
\end{figure}

Assuming all the BBH mergers observed by LIGO/Virgo during first and second observing runs \cite{LIGOScientific:2018mvr} are PBHs which originate from \eqref{eq:curvature_spectrum}, and by following the method in~\cite{Raidal:2018bbj}, we take a maximum likelihood analysis to find the best-fit parameters of our fitting formula of PBH mass function \eqref{eq:fmap}.
The log-likelihood function is
\begin{equation}\label{eq:likelihood}
    \mathcal{L} = \sum_{j} \ln \frac{\int P(m_{1},m_{2},z)\ p^{(j)}(m_{1}^{(j)}|m_{1})\ p^{(j)}(m_{2}^{(j)}|m_{2})\ p^{(j)}(z^{(j)}|z) \theta(\rho(m_{1},m_{2},z)-\rho_{c})}{\int P(m_{1},m_{2},z) \theta(\rho(m_{1},m_{2},z)-\rho_{c})},
\end{equation}
where
\begin{equation}
    dP(m_{1},m_{2},z) \propto dR (m_{1},m_{2},z) dV_{c}(z)
\end{equation}
is the probability density of having a BBH consisting of two black holes with mass $m_{1}$ and $m_{2}$ merging at redshift $z$, and $V_{c}(z)$ is the comoving volume.
The experimental uncertainties of black hole mass and redshift are accounted by $p^{(j)}(m^{(j)}|m)$ and $p^{(j)}(z^{j}|z)$, which denote the probability of observing black hole mass $m^{(j)}$ and redshift $z^{(j)}$ given that the true mass is $m$ and the true redshift is $z$.
$p^{(j)}$ is taken to be a Gaussian distribution with mean and variance given in \cite{Raidal:2017mfl}.
The detectability of detectors based on the signal-to-noise ratio (SNR) $\rho$ is implemented by step function $\theta$, and the detectability threshold is taken to be $\rho_{c}=8$. The results of log-likelihood $\mathcal{L}$ and normalized likelihood $L$ are shown in figure \ref{fig:logl}, and $\hat{m}=17.1_{-3.9}^{+5.1}$ at $99\%$ confidence level.

\begin{figure}[htpb]
    \centering
    \includegraphics[width=0.9\textwidth]{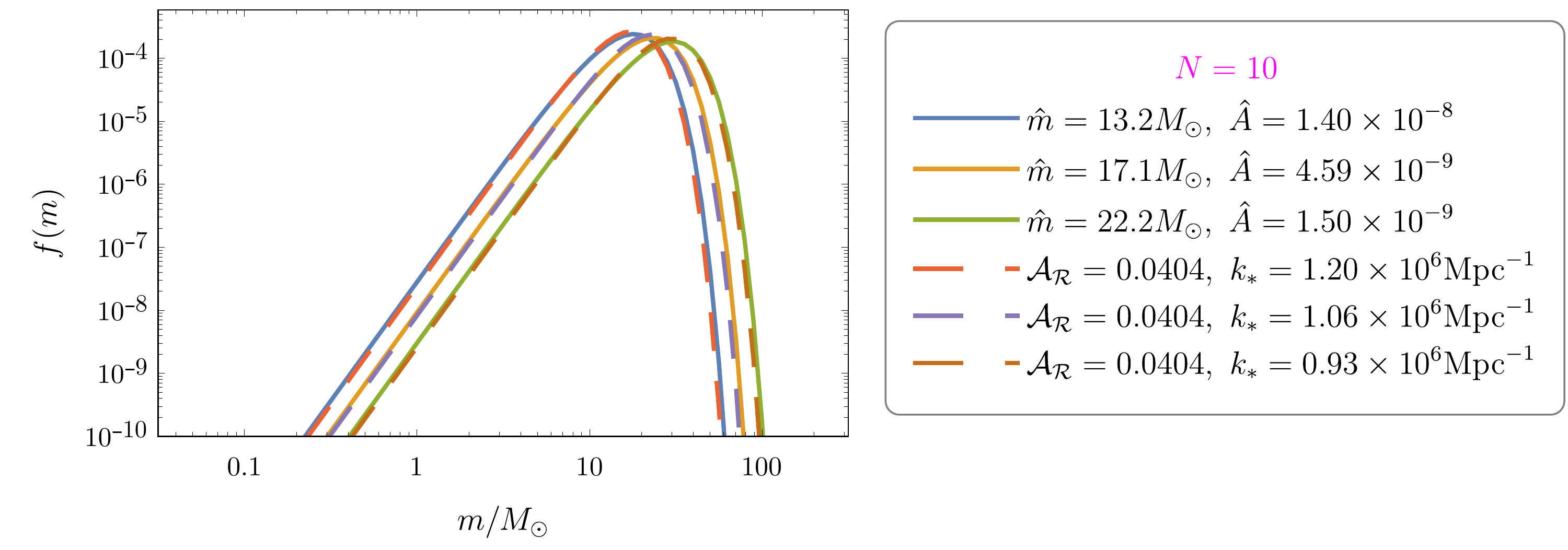}
    \caption{The parametrized and exact mass functions of PBHs with parameters fitting from $N=10$ LIGO detections are shown with solid and dashed lines respectively. Best-fit parameters are $\hat{m}=17.1 M_{\odot}$ and $\hat{A}=4.59\times 10^{-9}$, which can be converted into $\mathcal{A}_{\mathcal{R}}=0.0404$ and $k_{*}=1.06\times 10^{6} \Mpc^{-1}$.}
    \label{fig:N10d}
\end{figure}

Since the maximum likelihood analysis can only find the best fit of $\hat{m}$, we use the number of BBH merger events that LIGO observed to constraint $\hat{A}$.
The number of BBH merger events that LIGO should observe during period $\Delta t$ is
 \begin{equation}
     N = \Delta t \int dR(m_{1}, m_{2}, z) dV_{c}(z) \theta(\rho(m_{1}, m_{2}, z)- \rho_{c}).
\end{equation}
During the first and second runs $\Delta t \simeq 165$ days, LIGO has observed $N_{\mathrm{obs}}=10$ BBH merger events \cite{LIGOScientific:2018mvr}, which can be converted into the constraint of $\hat{A}$. The results are shown in figure \ref{fig:N10d}.
The best-fit parameters are $\hat{m}=17.1 M_{\odot}$ and $\hat{A}=4.59\times 10^{-9}$, which can be converted into $\mathcal{A}_{\mathcal{R}}=0.0404$ and $k_{*}=1.06\times 10^{6} \Mpc^{-1}$, and are compatible with the observational constraints to the fraction of PBHs in dark matter as is shown in figure \ref{fig:fpbh}.

\afterpage{
\begin{figure}[htpb]
    \centering
    \includegraphics[width=0.8\textwidth]{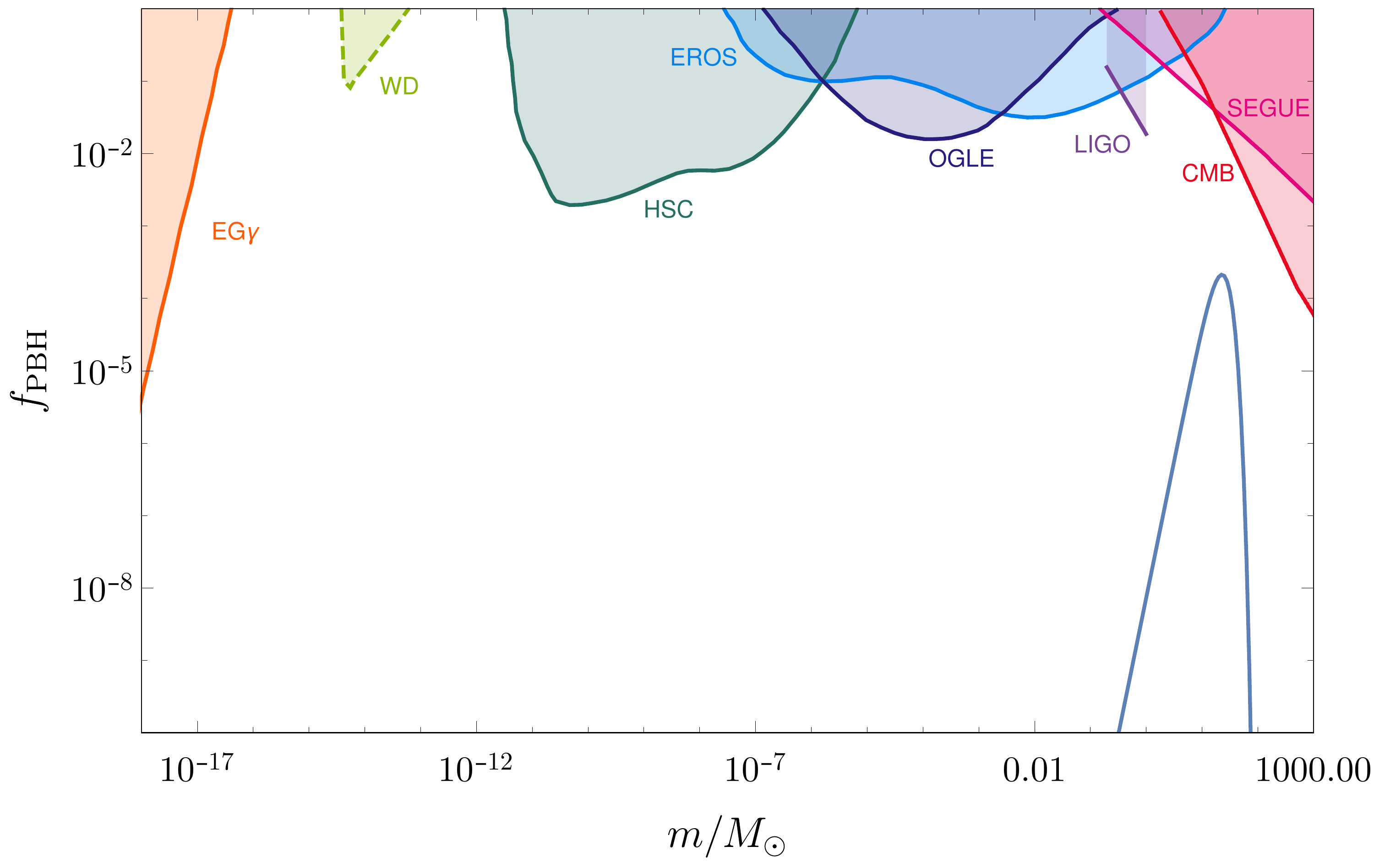}
    \caption[]{Best fitting of LIGO detections and observational constraints to the fraction of PBHs in dark matter.
        We adopt constraints from the extragalactic gamma ray background (EG$\gamma$ \cite{Carr:2009jm}), white dwarf explosions (WD \cite{Graham:2015apa}, see \cite{Montero-Camacho:2019jte} for criticism), gravitational lensing events (HSC \cite{Niikura:2017zjd}, EROS \cite{Tisserand:2006zx}, OGLE \cite{Niikura:2019kqi}), GWs (LIGO have not detected sub-solar-mass black holes~\cite{Authors:2019qbw}), dynamical effects (SEGUE \cite{Koushiappas:2017chw}), and cosmic microwave background (CMB~\cite{Poulin:2017bwe})\footnotemark.
}
\label{fig:fpbh}
\end{figure}
\footnotetext{As is discussed in~\cite{Poulin:2017bwe}, spherical accretion assumption is likely to break down and an accretion disk should form in the dark age, which will improve the CMB constraints on PBHs by two orders of magnitude. We appreciate the referee of JCAP for pointing out this. }
}

Given the best-fit parameters $\mathcal{A}_{\mathcal{R}}=0.0404$ and $k_{*}=1.06 \times 10^{6} \Mpc^{-1}$, we can calculate the spectrum of induced GWs by Eq.\eqref{eq:spectrum_GW} until equality. After the source term re-enters the horizon, it will decay rapidly, leaving the induced GWs evolving as free propagating radiation decaying as $a^{-4}$. The GW spectrum today can be connected to the equality by
\begin{equation}
    \begin{aligned}
        \Omega_{\mathrm{GW}}(k,\eta_{0}) &\equiv \frac{\rho_{\mathrm{GW}}(k,\eta_{0})}{\rho_{0}}
                                          = 2 \Omega_{r,0} \left( \frac{g_{*s}(T_{\mathrm{eq}})}{g_{*s}(T_{0})} \right)^{-\frac{4}{3}} \frac{g_{*r}(T_{\mathrm{eq}})}{g_{*r}(T_{0})} \Omega_{\mathrm{GW}}(k,\eta_{\mathrm{eq}}).
    \end{aligned}
\end{equation}
We plot $\Omega_{\mathrm{GW}}(k,\eta_{0})h^{2}$ in figure \ref{fig:OmegaGW} labelled by $F_\text{NL}=0$. It is easy to see that the best-fit configuration from LIGO detections is ruled out by the EPTA limits on stochastic GW background.
But we must emphasize that there are only 10 samples for the detection, which are not enough for statistics.
If more heavier BBHs are observed by LIGO in future, it is possible for the best-fit $k_{*}$ to be smaller, thus the peak of $\Omega_{\mathrm{GW}}(\eta_{0},k)h^{2}$ may be shifted to the left and be compatible with EPTA. 
In this paper, we will work under the assumption that, even if future detections of more LIGO/VIRGO events would not significantly change the currently observed merger rate, one could still be able to resolve this conflict simply by introducing some non-Gaussianity in the scalar perturbation, as we will show in the next section.

\section{Non-Gaussian curvature perturbation}\label{sec:NG}

The primordial non-Gaussianity of the curvature perturbation may come from the self coupling of the inflaton, the coupling of inflaton to other fields, or a non-Bunch-Davies initial condition. The non-Gaussianity may enhance or suppress the formation of PBHs, depending on the signature of the nonlinear parameter $f_\text{NL}$. Besides, the curvature perturbation is not the quantity we can observe directly today. The density contrast $\Delta$ that some authors use in the calculation of PBH formation has a nonlinear relation with the curvature perturbation $\mathcal{R}$ on scales comparable to the Hubble horizon, which becomes crucial at the horizon reentry for the PBH formation~\cite{Atal:2019cdz,Young:2019yug,DeLuca:2019qsy,Kawasaki:2019mbl}. All of these introduce non-Gaussianity in the curvature perturbation, which can have characterisitc impacts on the induced GWs ~\cite{Nakama:2016gzw,Garcia-Bellido:2016dkw,Cai:2018dig,Unal:2018yaa,Cai:2019amo,Yuan:2019udt}. It can be either a suppression or an enhancement on the peak value of the GW spectrum, and the former case can help to solve the conflict we encountered in the last section. 

Following \cite{Luo:1992er,Verde:1999ij,Verde:2000vr,Komatsu:2001rj,Bartolo:2004if,Boubekeur:2005fj,Byrnes:2007tm}, we consider a local-type non-Gaussian curvature perturbation $\mathcal{R}(\bm{x})$,
\begin{equation}\label{def:NG}
    \mathcal{R}(\bm{x}) = \mathcal{R}_{\mathrm{G}}(\bm{x}) + F_{\mathrm{NL}} \left[ \mathcal{R}_{\mathrm{G}}^{2}(\bm{x}) -\langle \mathcal{R}_{\mathrm{G}}^{2}(\bm{x}) \rangle \right] ,
\end{equation}
where $F_\text{NL}$ is the nonlinear parameter for the curvature perturbation, relating to the nonlinear parameter $f_\text{NL}$ for the Newton potential $\Phi$ by $F_\text{NL}=(3/5)f_\text{NL}$. We will see that only positive $F_\text{NL}$ is useful for suppressing the GW spectrum, because it will increase the PBH abundance given the same variance. $\mathcal{R}_{\mathrm{G}}(\bm{x})$ is the Gaussian curvature perturbation, which can be solved from \eqref{def:NG} as~\cite{Byrnes:2012yx,Young:2013oia}
\begin{equation}
    \mathcal{R}_{\mathrm{G}\pm} (\mathcal{R}) = (2 F_\text{NL})^{-1} \left( -1 \pm \sqrt{1+4 F_\text{NL} \mathcal{R}+4 F_\text{NL}^{2} \langle \mathcal{R}_{\mathrm{G}}^{2} \rangle} \right) .
\end{equation}
The probability distribution function of this full curvature perturbation $\mathcal{R}$ can be derived by requiring $\mathcal{R}_{\mathrm{G}}$ obeying the Gaussian distribution,
\begin{equation}
P_{\mathrm{G}}(\mathcal{R}_{\mathrm{G}\pm}) = \frac{1}{\sqrt{2\pi \langle \mathcal{R}_{\mathrm{G}}^{2} \rangle}} \exp \left( -\frac{\mathcal{R}_{\mathrm{G}\pm}^{2}}{2\langle \mathcal{R}_{\mathrm{G}}^{2} \rangle} \right),
\end{equation}
with the Jacobian
\begin{equation}
    P(\mathcal{R}) d\mathcal{R} = \sum_{i=+,-} \left| \frac{d \mathcal{R}_{\mathrm{G},i}}{d \mathcal{R}} \right| P_{\mathrm{G}}(\mathcal{R}_{\mathrm{G},i}) d \mathcal{R}.
\end{equation}
The initial PBH mass fraction can be written as
\begin{equation}
    \begin{aligned}
        \beta &\simeq 2 \int_{\mathcal{R}_{c}}^{\infty} P(\mathcal{R}) d\mathcal{R}
    = \erfc \left( \frac{\mathcal{R}_{\mathrm{G}+}(\mathcal{R}_{c})}{\sqrt{2 \langle \mathcal{R}_{\mathrm{G}}^{2} \rangle}} \right) + \erfc \left( -\frac{\mathcal{R}_{\mathrm{G}-}(\mathcal{R}_{c})}{\sqrt{2 \langle \mathcal{R}_{\mathrm{G}}^{2} \rangle}} \right),
    \end{aligned}
\end{equation}
where $\mathcal{R}_{c} \simeq \frac{9}{2\sqrt{2}} \Delta_{c} \simeq 1.4$ is the critical curvature perturbation for PBH formation \cite{Young:2014ana}, and $\langle \mathcal{R}_{\mathrm{G}}^{2} \rangle\simeq \mathcal{A}_{\mathcal{R}} $ for $\sigma_{*}\ll k_{*}$.

\begin{figure}[htpb]
    \centering
    \includegraphics[width=0.5\textwidth]{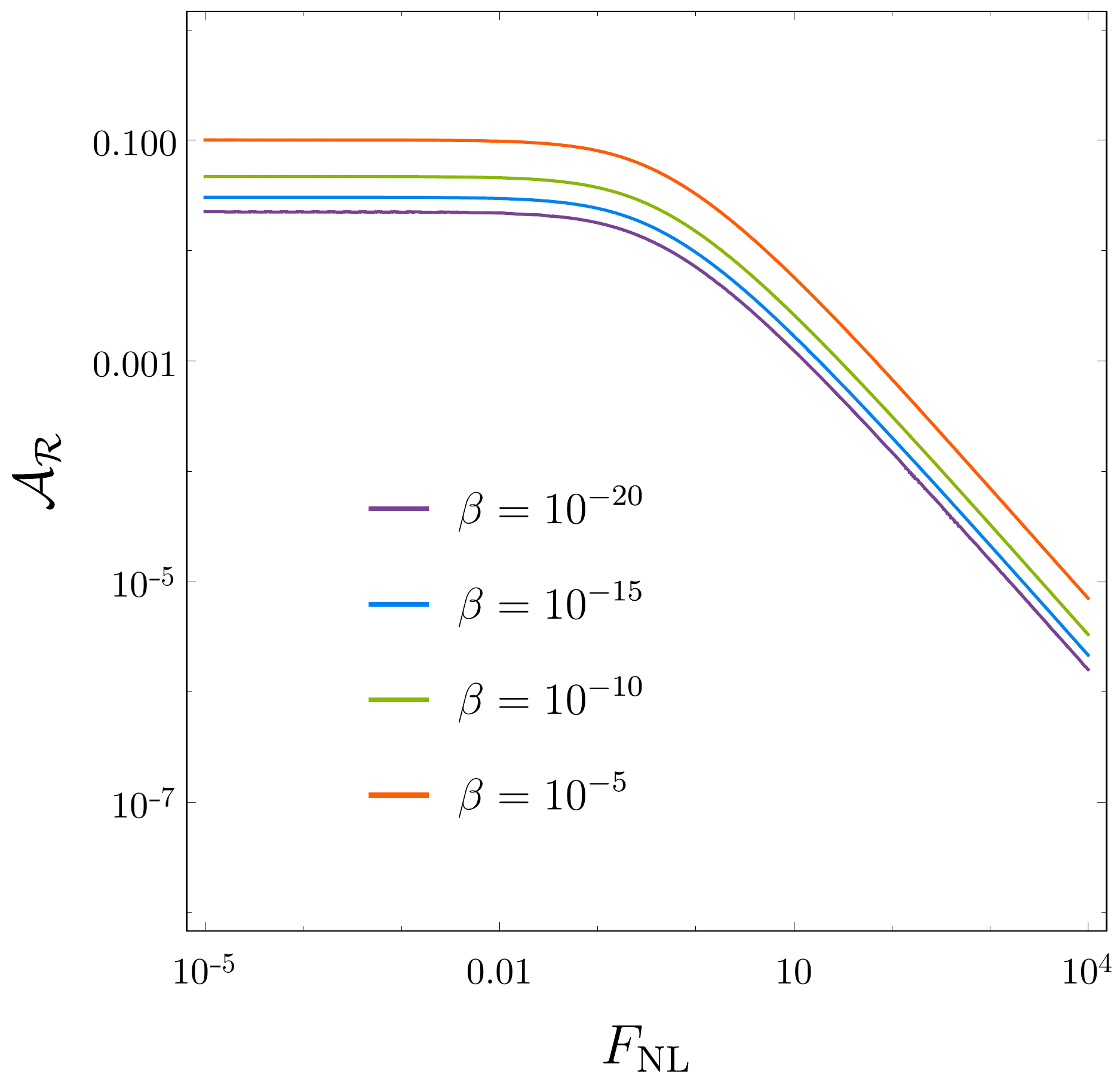}
    \caption{Initial PBH mass fraction $\beta$ as a function of $F_\text{NL}$ and $\mathcal{A}_{\mathcal{R}}$.}
    \label{fig:ARfNLbeta}
\end{figure}

In figure \ref{fig:ARfNLbeta}, $\mathcal{A}_\mathcal{R}$ with different $\beta$ as a function of $F_\text{NL}$ are shown.
Given a fixed $\beta$ from observations, we see that $F_\text{NL} \mathcal{A}_{\mathcal{R}}$ is constant while $F_\text{NL} \rightarrow \infty$,
\begin{equation}
        A_{0} \equiv \lim_{F_\text{NL}\rightarrow 0}\mathcal{A}_{\mathcal{R}}= \frac{\mathcal{R}_{c}^{2}}{2\ \Ierfc(\beta)^{2}},
\end{equation}
\begin{equation}
  A_{\infty} \equiv \lim_{F_\text{NL}\rightarrow \infty} F_\text{NL} \mathcal{A}_{\mathcal{R}} = \frac{\mathcal{R}_{c}}{2\ \Ierfc(\beta/2)^{2}-1},
\end{equation}
where $\Ierfc(x)$ is the inverse function of $\erfc(x)$. For $\beta \ll 1$, we can approximate
 \begin{equation}
    A_{0} \simeq \mathcal{R}_{c} A_{\infty}.
\end{equation}

\begin{figure}[htpb]
    \centering
    \includegraphics[width=0.8\textwidth]{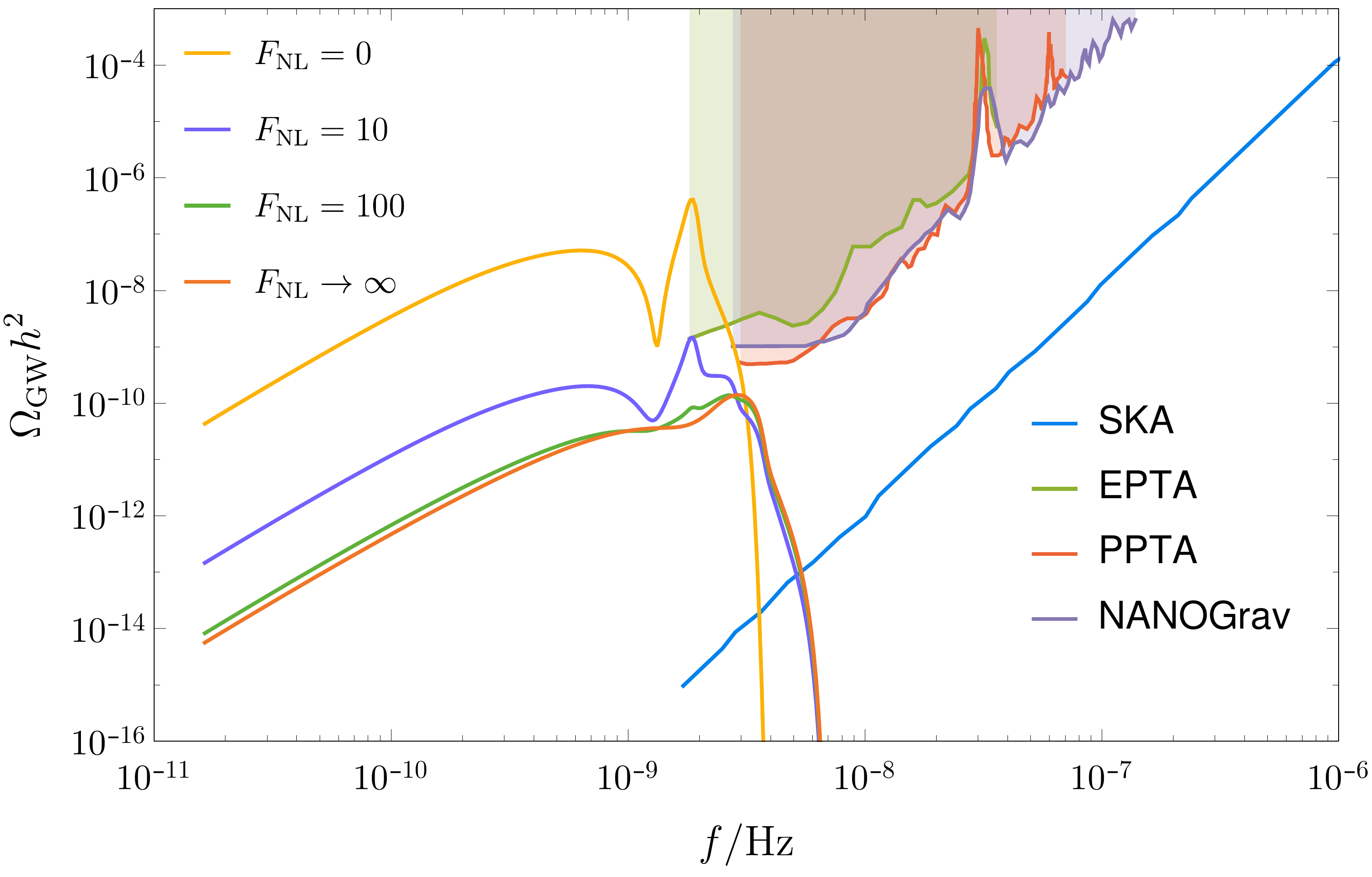}
    \caption{The GW spectrum with $F_\text{NL}=0,~10,~100$ and $F_\text{NL}\rightarrow \infty$ fit from LIGO detections with respect to the sensitivities of current/future PTA projects.
    The current constraints (shaded) are given by EPTA \cite{Lentati:2015qwp}, PPTA \cite{Shannon:2015ect}, NANOGrav\cite{Arzoumanian:2018saf}, and the future sensitivity curve of SKA is depicted following \cite{Moore:2014lga}.	
}
    \label{fig:OmegaGW}
\end{figure}

The GW spectrum from non-Gaussian curvature perturbations is given by \cite{Cai:2019amo,Cai:2018dig}
\begin{align}\label{eq:GWNG}
\Omega_\mathrm{GW}(k)=\frac{1}{54}\int_0^\infty\mathrm{d}v\int_{|1-v|}^{1+v}\mathrm{d}u\mathcal{T}(u,v)\mathcal{P}_{\mathcal{R}}^\mathrm{NG}(ku)\mathcal{P}_{\mathcal{R}}^\mathrm{NG}(kv),
\end{align}
where
\begin{equation}
    \begin{aligned}
        \mathcal{P}_{\mathcal{R}}^\mathrm{NG}(k)&=\mathcal{P}_{\mathcal{R}}(k)+F_\mathrm{NL}^2\int_0^\infty\mathrm{d}v\int_{|1-v|}^{1+v}\mathrm{d}u\frac{1}{u^2v^2}\mathcal{P}_{\mathcal{R}}(ku)\mathcal{P}_{\mathcal{R}}(kv)\\
&=\frac{\mathcal{A}_{\mathcal{R}} k_{*}}{\sqrt{2\pi}\sigma_{*}}\exp \left[ - \frac{(k-k_{*})^{2}}{2\sigma_{*}^{2}} \right] + F_\text{NL}^{2} \mathcal{A}_{\mathcal{R}}^{2} \frac{k^{2}}{k_{*}^{2}}\ \frac{1}{2} \erf(\frac{k}{2\sigma_{*}}) \left[ 1+\erf(\frac{2k_{*}-k}{2\sigma_{*}}) \right],
    \end{aligned}
\end{equation}
for $\sigma_{*}\ll k_{*}$. Roughly speaking, Eq.\eqref{eq:GWNG} can be estimated as
\begin{equation}
    \Omega_{\mathrm{GW}}(k)\simeq \mathcal{A}_{\mathcal{R}}^{2} \Omega^{(0)}(k) + F_\text{NL}^{2} \mathcal{A}_{\mathcal{R}}^{3} \Omega^{(2)}(k) + F_\text{NL}^{4} \mathcal{A}_{\mathcal{R}}^{4} \Omega^{(4)}(k),
\end{equation}
where $\Omega^{(0)}(k)$, $\Omega^{(2)}(k)$, $\Omega^{(4)}(k)$ are the corresponding integral terms, and they are all of order $\mathcal{O}(1)$ at the peak. Therefore, fixing the PBH abundance $\beta$ that is determined by our fit, the GW spectrum induced by the Gaussian part of the curvature perturbation is $\Omega_\text{GW}^{(0)}\sim A_0^2$, while the extremely non-Gaussian part of GW spectrum is $\Omega_{\mathrm{GW}}^{f_\text{NL}\rightarrow\infty}\sim\mathcal{O}(A_\infty^4)$. This gives
\begin{equation}
    \frac{\Omega_{\mathrm{GW}}^{F_\text{NL}\rightarrow\infty}(k_{*})}{\Omega_{\mathrm{GW}}^{(0)}(k_{*})} \sim \mathcal{O}\left(\frac{A_{0}^{2}}{\mathcal{R}_{c}^{4}}\right) \sim \mathcal{O}(10^{-4})
\end{equation}
with $A_{0}=0.0404$ obtained from the best-fit parameter in section \ref{sec:fit}.
Therefore, we can see that the peak value of the GW spectrum is greatly suppressed, which makes it possible to evade the constraint from EPTA. 
Accurate calculations for $\Omega_\text{GW}$ from \eqref{eq:GWNG} is done and depicted in Fig.~\ref{fig:OmegaGW} for a fixed PBH abundance $\beta=3.4\times 10^{-12}$ obtained from Eq.\eqref{eq:beta}, for $F_\text{NL}=0,~10,~100$ and $F_\text{NL}\rightarrow \infty$.
It is shown explicitly that all black holes observed by LIGO can be PBHs, if the curvature perturbation is non-Gaussian with $F_\text{NL}\gtrsim 10$.

Here we would like to mention that the future radio telescope project Square Kilometer Array (SKA)~\cite{Janssen:2014dka} can extend the detecting accuracy to $\Omega_{\mathrm{GW}}\lesssim10^{-14}$ at around $4~\text{nHz}$. We depict the expected SKA sensitivity curve in Fig.~\ref{fig:OmegaGW}. It is easy to see that whether the LIGO/VIRGO detection events are mostly consists of PBHs can be easily checked by SKA. We can also see that as the non-Gaussian peak has a higher frequency, we do not need 20 years to see that peak. 5 to 10 years of observation will be enough to see whether the non-Gaussian peak in GW spectrum exists or not.

\section{Conclusion and Discussions}\label{sec:con}

The possibility that LIGO-detected black holes may be PBHs has attracted much attention since the PBH merger rate fits that of the LIGO estimation~\cite{Sasaki:2016jop,Bird:2016dcv}.  These PBHs can be distinguished with black holes formed by astrophysical processes by studying their spin distribution~\cite{Chiba:2017rvs,Mirbabayi:2019uph,DeLuca:2019buf,Fernandez:2019kyb,He:2019cdb}, redshift dependence of the merger rate~\cite{Nakamura:2016hna,Chen:2019irf}, mass dependence of the merger rate~\cite{Kocsis:2017yty,Liu:2019rnx}, and eccentricity~\cite{OLeary:2008myb,Cholis:2016kqi}. Currently there is no clear evidence for the primordial or astrophysical nature of the BBHs, yet there is some tension in the PTA constraints on stochastic GWs accompanied with the PBHs.

Solar-mass PBHs form substantially if the primordial curvature perturbation has a peak on comoving scale of $0.1~\text{pc}$, which also induces a stochastic background of GWs with its peak frequency within the reaches of PTA observations. Therefore, currently null detection of such GW background put a conservative bound on the peak amplitude of curvature perturbation and hence on the mass function of PBHs, which is in a seemingly mild tension with the best-fit parameters obtained by the maximum likelihood analysis of merger rate from the LIGO/VIRGO O1 and O2 events. To resolve this potential tension, we consider the GWs induced by curvature perturbations with local-type non-Gaussianity, and find that current constraints on the stochastic background of GWs from PTA observations could be evaded, if the nonlinear parameter is positive and larger than $\mathcal{O}(10)$. 

We must emphasize that the statistic error may be large as the current number of samples given by the LIGO/VIRGO O1 and O2 events is limited. Also, it may be possible that some of the events are not primordial thus the induced GWs are overestimated. More merger observations are needed to determine the nature of the black holes and their mass function, which will be provided by the third run of LIGO/VIRGO and the upcoming KAGRA~\cite{Somiya:2011np} and LIGO-India~\cite{Unnikrishnan:2013qwa}, and the future space-based experiments like LISA \cite{AmaroSeoane:2012km,AmaroSeoane:2012je,Audley:2017drz}, Taiji~\cite{Guo:2018npi}, Tianqin~\cite{Luo:2015ght}, BBO~\cite{Crowder:2005nr,Corbin:2005ny} 
and DECIGO~\cite{Kawamura:2006up,Kawamura:2011zz}. Our paper provide a possible solution to the tension if it still exists in the future statistics with more events, and we expect SKA can verify or falsify the primordial nature of the binary black holes, which is independent of the other criteria. 

\section*{Acknowledgement}
We thank Misao Sasaki and Ying-li Zhang for useful discussions.
RGC and XYY are supported by the National Natural Science Foundation of China Grants Nos.11435006, 11647601, 11690022, 11821505, 11851302, and by the Strategic Priority Research Program of CAS Grant NO. XDB23010500 and No.XDB23030100, and by the Key Research Program of Frontier Sciences of CAS.
SP is supported by the MEXT/JSPS KAKENHI No. 15H05888 and the World Premier International Research Center Initiative (WPI Initiative), MEXT, Japan.
SJW is supported by the postdoctoral scholarship of Tufts University.

\bibliographystyle{JHEP}
\bibliography{ref}

\end{document}